# Forces and energetics of intermittent swimming

Daniel Floryan · Tyler Van Buren · Alexander J. Smits

**Abstract** Experiments are reported on intermittent swimming motions. Water tunnel experiments on a nominally two-dimensional pitching foil show that the mean thrust and power scale linearly with the duty cycle, from a value of 0.2 all the way up to continuous motions, indicating that individual bursts of activity in intermittent motions are independent of each other. This conclusion is corroborated by PIV flow visualizations, which show that the main vortical structures in the wake do not change with duty cycle. The experimental data also demonstrate that intermittent motions are generally energetically advantageous over continuous motions. When metabolic energy losses are taken into account, this conclusion is maintained for metabolic power fractions less than 1.

**Keywords**   unsteady propulsion · burst and coast · bio-inspired

## 1 Introduction

Many aquatic animals, such as large sharks and seals [1] to small schooling fish [2], exhibit an intermittent swimming behavior, sometimes called burst-and-coast swimming. Fish practice intermittent swimming while hunting, fleeing a predator, pursuing a mate, or while starving, and exhibit a wide range of ratios of burst to coast times [3]. Our primary interest here is to examine the potential energy benefit of intermittent swimming in cruising conditions.

In this respect, Weihs [4] developed a simple theoretical model and showed that fish could achieve up to 50% savings in energy through intermittent swimming, which was supported by observations on the swimming range of salmon. Videler and Weihs [5] then found that cod and saithe choose minimum and maximum swimming velocities that match predicted theoretical optima.

The key parameter that differentiates intermittent from continuous swimming in Weihs' model is $\alpha$, the ratio of drag during active swimming to drag during coasting. The model showed that intermittent swimming is potentially energetically advantageous only for values of $\alpha > 1$. Lighthill [6] and Webb [7] noted that the total drag on fish is 3-5 times higher when they are actively swimming than when they do not actuate their bodies, which corresponds to a range of $\alpha$ where intermittent swimming is expected to be energetically advantageous. Lighthill suggested that the drag increases during active swimming due to the thinning of the boundary layer, and so boundary layer thinning may be hypothesized as the mechanism responsible for the energetic benefits of intermittent swimming; we emphasize that this is a viscous mechanism. As the author himself noted, however, this hypothesis was tentative and untested, and more recent work suggests that boundary layer thinning during unsteady swimming motions could lead to only about a 90% increase in drag [8].

Similar simplified hydromechanic approaches were followed by Blake [9] and Chung [10], but the main focus of their work was on the impact of the shape of the body of the fish and how to minimize drag, factors that directly impact the energy expenditure of intermittent motions. Numerical simulations by Wu et al. [11] show similar energetic benefits as the theory.

To understand the consequences of intermittent swimming, we have conducted an experimental investigation of intermittent propulsion by a rigid pitching foil. The performance of pitching and heaving foils in continuous motion is relatively well understood, and Floryan et al. [12] recently presented a comprehensive scaling argument that collapses the available data on thrust and power well. In that context, our aim is to understand how intermittent swimming motions affect the forces produced, and the energy expended, and to develop a scaling analysis to describe these changes. We seek to answer three specific questions: (1) do intermittent

The project was supported by the US Office of Naval Research, under N00014-14-1-0533.

D. Floryan (✉) · T. Van Buren · A. J. Smits
Department of Mechanical and Aerospace Engineering, Princeton University, Princeton, NJ 08544, USA
e-mail: dfloryan@princeton.edu

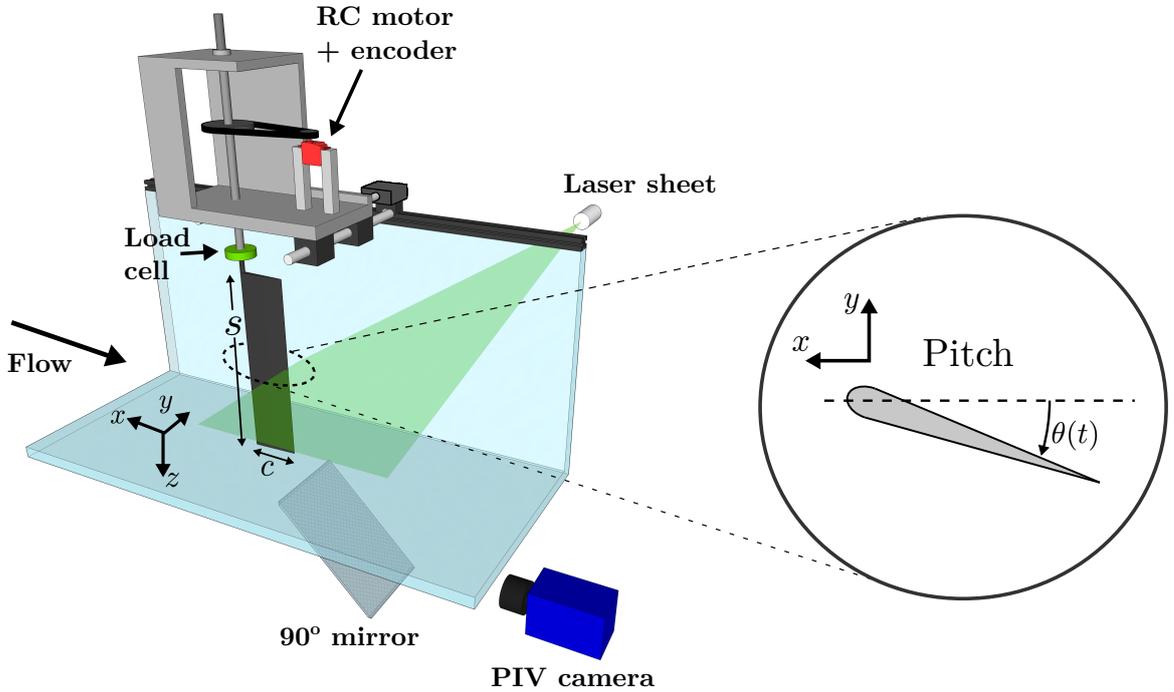

Figure 1: Experimental setup.

motions reduce the energy needed to traverse a given distance if metabolic losses are ignored, (2) how does this conclusion change when metabolic energy losses are taken into account, and (3) do intermittent motions reduce the energy needed to traverse a given distance when the average speed is fixed?

## 2 Experimental setup

Experiments were conducted using a rigid pitching foil in a recirculating free-surface water channel, as shown in figure 1. The water channel test section is 0.46 m wide, 0.3 m deep, and 2.44 m long. Baffles upstream and downstream of the foil minimized surface waves. The free-stream velocity, $U_\infty$, was fixed at 100 mm/s for performance testing and 60 mm/s for wake measurements.

The propulsor was a two-dimensional teardrop foil with a chord of $c$ = 80 mm, maximum thickness of 8 mm, and span of $s$ = 279 mm. The performance was measured by a six component force/torque sensor (ATI Mini40), with force and torque resolutions of $5 \times 10^{-3}$ N and $1.25 \times 10^{-4}$ N·m, respectively. The pitching motions were generated by an RC motor (Hitec HS-8370TH) monitored via a separate encoder. Pitching motions were sinusoidal, and the amplitudes were varied from $\theta_0$ = 5° to 15° every 5°. For intermittent motions, duty cycles of $\Delta$ = 0.2 to 0.9 every 0.1 were explored and compared to continuous motion ($\Delta$ = 1). A duty cycle of 0.2, for example, means that the foil completes one full period of pitching, and then stops moving with $\theta$ = 0 for a time equal to four periods. The actuation frequency, that is, the frequency of the active portion of the cycle, varied from $f$ = 0.2 to 1.5 Hz every 0.1 Hz. This yielded a Strouhal number, $St = 2fc\,sin(\theta_0)/U_\infty$, range from 0.05 to 0.4. Each trial consisted of 30 cycles, and the data were averaged over the middle 20 cycles. Each trial was repeated a minimum of 3 times to ensure repeatability and reduce uncertainty.

The wake velocity measurements were taken at the center-span of the foil with particle image velocimetry (PIV). Silver coated hollow ceramic spheres (Potter Industries Inc. Conduct-O-Fil AGSL150 TRD) were used to seed the flow, illuminated by a CW argon-ion laser (Spectra Physics 2020). An 8-bit monochrome CCD camera (MotionXtra HG-LE) with 1128×752 resolution was used to acquire images at 25 Hz. Images were processed sequentially with commercial DaVis software using spatial correlation interrogation window sizes of 64×64 and twice at 32×32 with 50% overlap. The (cropped) measurement region covered 86 mm in the streamwise direction and 84 mm across, with a resolution of 7 vectors per 10 mm. Average and instantaneous velocity errors are estimated to be 2.7% and 1-5%, respectively [13].

## 3 Forces and power

We begin by considering experiments in which the foil is fixed in the streamwise direction and pitched sinusoidally either continuously or intermittently. The propulsive performance is described using the conventional definitions of thrust and power coefficients, where

$$C_T = \frac{F_x}{\frac{1}{2}\rho U_\infty^2 sc}, \qquad C_P = \frac{M_z \dot{\theta}}{\frac{1}{2}\rho U_\infty^3 sc}. \qquad (1)$$

Here, $F_x$ is the streamwise force, $M_z$ is the spanwise torque, $\dot{\theta}$ is the angular velocity of pitching, and $\rho$ is the fluid density.

In switching from swimming to non-swimming, the form of the transition is important since it has an impact on both the thrust and power, as shown in figure 2. We therefore define a smoothing parameter, $\xi$, where $\xi = 0$ is unsmoothed (see inlay of figure 2). The waveforms are smoothed via a Gaussian filter, where $\xi$ is the width of the Gaussian kernel relative to the width of the active portion of the wave. As the waveform becomes smoother, the thrust and power decrease because the starting and stopping accelerations decrease. To make comparisons meaningful, we apply a smoothing parameter of $\xi = 0.1$ for all experiments.

The thrust and power coefficients are shown in figure 3 for a range of duty cycles and pitch amplitudes. Both the thrust and power increase nonlinearly with Strouhal number and duty cycle in all cases. In figure 4 we show the same results scaled with the duty cycle, which collapses all of the data points onto a single curve. For intermittent motions, it appears to be sufficient to correct the time-averaged force such that the averaging is only done on the active portion of the cycle. This procedure is similar to that used by Akoz & Moored [14], who scale the thrust for intermittent motions with the frequency of motion, which effectively accounts for duty cycle. This result has the important implication that the thrust and power generated in each actuation cycle can be treated independently. The whole is, indeed, the sum of its parts.

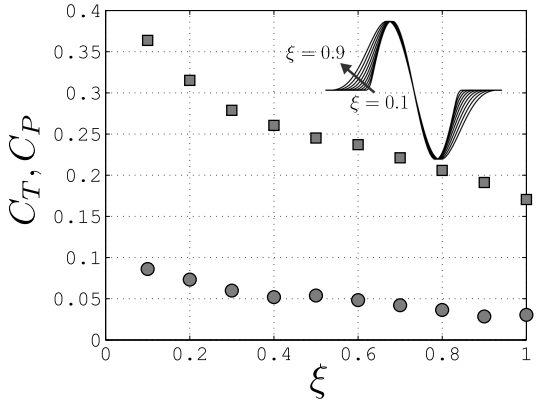

Figure 2: Time-averaged thrust (circular symbols) and power coefficients (square symbols) as they vary with smoothing parameter $\xi$ ($\theta_0 = 10°$, $f = 1$ Hz, $\Delta = 0.5$.)

Figure 5 shows the instantaneous wake vortex structure at the point in the generation cycle where the primary positive vortex is generated (labeled 1 in each figure). The continuous sinusoidal motion produces a typical thrust-producing "reverse von Kármán vortex street," generating two opposite-sense vortices per actuation cycle. The intermittent motion, however, produces two primary vortices (1 & 2) and two smaller secondary vortices (1' & 2'). These secondary vortices are due to the rapid start and stop of the intermittent motion. Similar vortex formations have been shown numerically for intermittent motions [11] and secondary vortices are also found in rapid starting and stopping square wave motions [15].

Changing the duty cycle appears to have little effect on the location and strength of the vortices produced per cycle, suggesting that there is minimal interaction between separate cycles. This finding is consistent with the observed scaling of the thrust and power coefficients with the duty cycle.

## 4 Free swimming performance

We now consider a free swimmer, no longer constrained to move at a fixed speed. Its motion is governed by Newton's Second Law

$$m\dot{u} = T - D, \qquad (2)$$

where $m$ is the mass, $u$ is the speed, $T$ is the thrust, and $D$ is the drag. We make the assumptions that the swimmer can be effectively split into a drag-producing part (the body) and a thrust-producing part (the propulsor), as illustrated in figure 6, and that these parts are independent of each other.

We further assume a quadratic drag law [16], such that

$$D = \frac{1}{2}\rho u^2 A_w C_D, \qquad (3)$$

where $\rho$ is the fluid density, $A_w$ is the wetted area of the body, and $C_D$ is the drag coefficient. We know from [12] that the thrust produced by a pitching foil is independent of the free-stream velocity, and so we can assume that the thrust measured from the fixed velocity experiments is the same as what would be measured if the foil were allowed to move. Furthermore, the appropriate velocity scale for a pitching foil is $fA$, where $f$ is the frequency of the motion, and $A$ is the amplitude of the motion. The non-dimensionalized governing equation is then

$$2\frac{m^*}{A^*}\frac{du^*}{dt^*} = C_T^* - C_D A_w^* u^{*2}, \qquad (4)$$

where $m^* = m/\rho c^2 s$ is the ratio of the body mass to the propulsor added mass, $A^* = A/c$ is the ratio of the amplitude to the chord, $u^* = u/fA$ is the non-dimensional speed, $t^* = tf$ is the non-dimensional time, $C_T^* = T/\frac{1}{2}\rho(fA)^2 sc$ is the thrust coefficient using the new velocity scale, $C_D$ is the drag coefficient as in (3), and $A_w^* = A_w/sc$ is the ratio of the body's wetted area to the propulsor's wetted area. The prefactor $m^*/A^*$ is the ratio of the body's mass to the mass of the fluid displaced by the propulsor.

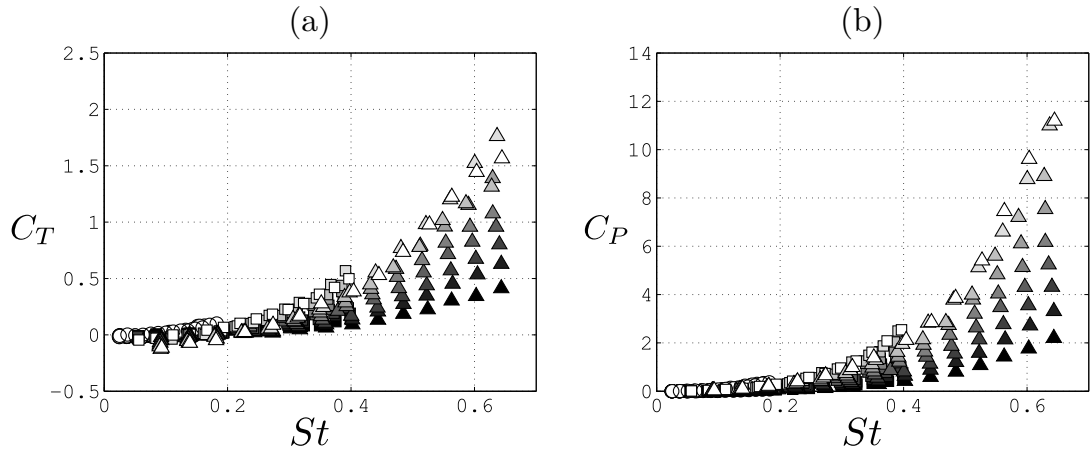

Figure 3: Time-averaged (a) thrust and (b) power coefficients as functions of Strouhal number. Dark to light symbols represent increasing duty cycles, ranging from $\Delta$ = 0.2 to 1 every 0.1. Symbols identify pitch amplitudes of $\theta_0 = 5°$ (circle), $10°$ (square), and $15°$ (triangle).

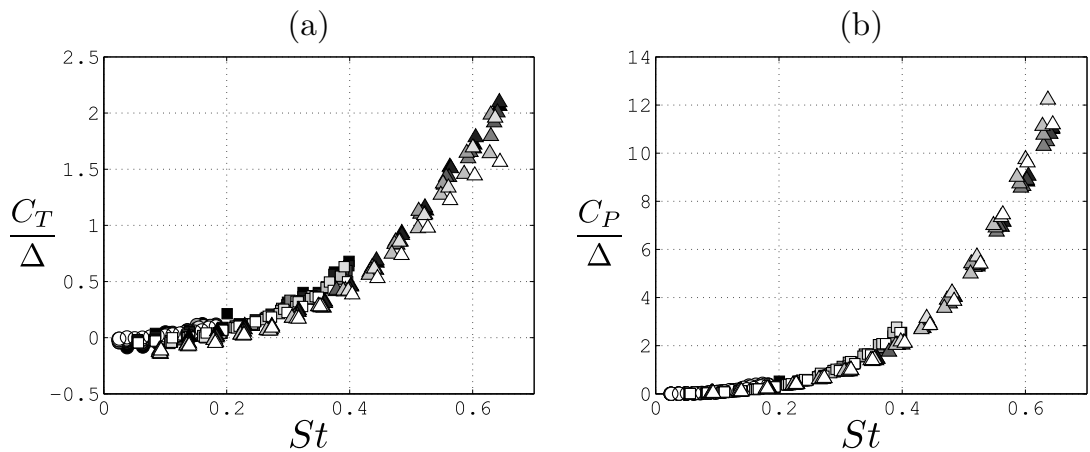

Figure 4: Time-averaged (a) thrust and (b) power coefficients normalized by duty cycle as functions of Strouhal number. Symbols and tones as in figure 3.

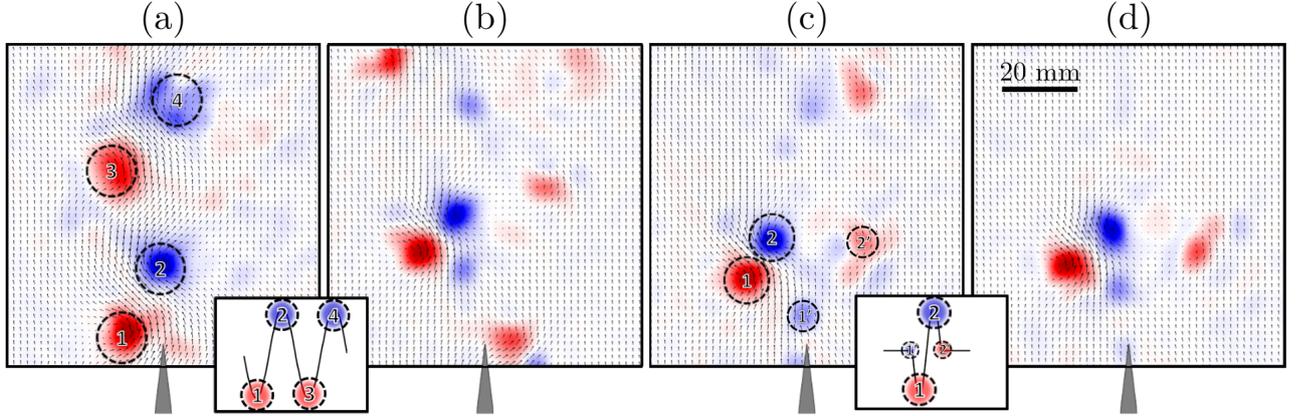

Figure 5: Instantaneous wake vorticity with in-plane velocity arrows overlaid for (a) continuous motion, (b) duty cycle of $\Delta = 0.25$, (c) $\Delta = 0.5$, and (d) $\Delta = 0.75$ and a Strouhal number of $St = 0.4$. Small graphics in the bottom right show the pitch angle over time and the generation points of each vortex in the the actuation cycle.

We find the mean speed for the free swimmer by inputting the phase-averaged thrust measured in our experiments into (4) and solving numerically. We solve using MATLAB's built-in integrator ODE45 [17, 18], evolving the system forward until it settles on the limit cycle. The drag coefficient is kept constant at 0.01 [19], $A^*$ varies with the data, and we vary the mass and area ratios over some orders of magnitude.

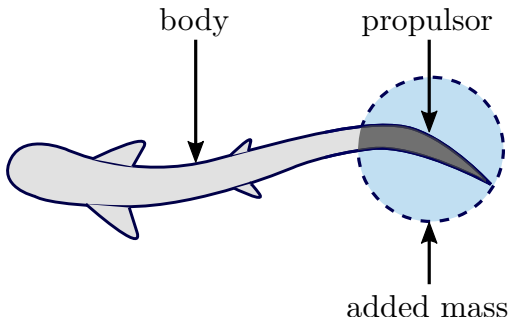

Figure 6: A swimmer can be simply represented as a combination of a drag-producing body and a thrust-producing propulsor.

4.1 Results on mean speed

We first compare the calculated mean speed $U^*_{mean}$ to the "steady-state" speed $U^*_{steady}$, that is, the speed found by setting $du^*/dt^* = 0$ in (4) and equating mean thrust to mean drag. The results, plotted in figure 7, show that the two speeds are nearly equal. The calculated mean speed falls below the steady value only for small values of the mass ratio $m^*$ and large values of the area ratio $A^*_w$, which represent unphysical realizations. Points with mean speed greater than the steady speed seem to be spurious.

The relation between the calculated mean and steady speeds can be understood by considering the problem in the frequency domain. Let

$$C_T^* = \sum_n C_n e^{i\omega nt}, \quad u^* = \sum_n u_n e^{i\omega nt} \quad (5)$$

be the Fourier series representing the thrust and speed, respectively. Since the thrust and speed are real, we require that $C_{-n} = \overline{C_n}$ and $u_{-n} = \overline{u_n}$, where the overbar denotes the complex conjugate. The equation for the zeroth Fourier mode of speed is then

$$u_0^2 = \frac{C_0}{C_D A_w^*} - \sum_{n \neq 0} u_n \overline{u_n}. \quad (6)$$

We see that the higher harmonics will act to decrease the mean speed. The effects of the mean thrust, drag coefficient, and area ratio on the mean speed are as expected.

To understand the effect of the mass ratio, we linearize (4) about the steady speed and take the Laplace transform. The system has a single left-half-plane pole at $-bC_D A_w^* A^*/m^*$, where $b$ is a constant arising from linearization. A greater mass ratio moves the pole closer to the origin, thus attenuating high frequencies. Intuitively, a greater mass ratio corresponds to a swimmer with greater inertia; the greater inertia will cause the speed to fluctuate less about its mean. According to (6), the attenuation of high frequencies with greater mass ratio brings the mean speed closer to the steady speed, explaining the observed behavior.

The effect of the area ratio on the mean speed can be understood in the same way. A smaller area ratio moves the pole closer to the origin, attenuating high frequencies. The mean speed should thus be closer to the steady speed as the area ratio is decreased.

We end by noting that the mean speed only deviates from the steady speed for extreme values of the parameters, unlikely to be encountered in nature, and for low values of the mean speed where experimental errors in our force measurements are relatively larger. For a wide range of the parameters, estimating the mean speed to be the steady speed appears to be a reasonable approximation.

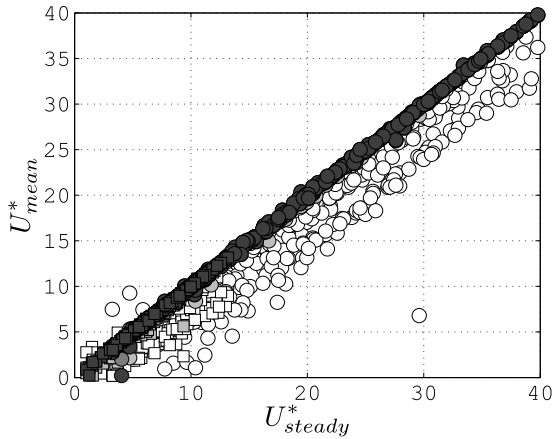

Figure 7: Mean speed versus steady speed for all frequencies, amplitudes, and duty cycles. Mass ratios are $m^* = 0.01, 0.1, 1, 10$ (light to dark symbols), and area ratios are $A_w^* = 1$ (circles) and $10$ (squares).

4.2 Results on energetics

Consider now the energetics of intermittent motions. Given a set of actuation parameters (frequency and amplitude), we would like to know whether changing the duty cycle decreases the energy required to travel a certain distance. This is captured by $\phi = E/E_0$, the ratio of the energy expended by intermittent motion to the energy expended by continuous motion with the same actuation. This energy ratio can be otherwise expressed as

$$\phi = \frac{E}{E_0} = \frac{Pt}{P_0 t_0} = \frac{Pd/U}{P_0 d_0 / U_0} = \frac{P U_0}{P_0 U},$$

where $P$ is the mean power, and $t$ is the time taken to travel a distance $d$ at a mean speed $U$. For the problem considered here, $d = d_0$. Values of $\phi < 1$ indicate that intermittent motions are energetically favorable, whereas values of $\phi > 1$ indicate that intermittent motions are unfavorable.

The results are plotted in figure 8. We only use data with $m^* = 1$ and $A_w^* = 10$, as trends should not vary with the mass and area ratios for reasonable values, as found earlier. We see that intermittent motions are almost always energetically favorable compared to continuous motions, with greater energy savings with decreasing duty cycle. We know from section 3 that $T \sim T_0 \Delta$ ($\Delta$ is the duty cycle), where $T$ is the mean thrust, and the symbol $\sim$ means "proportional to." Similarly, $P \sim P_0 \Delta$. Furthermore, figure 7 shows that $U \sim \sqrt{T}$ for a wide range of parameters. The energy ratio should then behave according to

$$\phi = \frac{P}{P_0} \frac{U_0}{U} \sim \Delta \frac{1}{\sqrt{\Delta}} = \sqrt{\Delta}.$$

Figure 8 indicates that the data follow this trend, which is not unexpected: a swimmer in this scenario does not expend any energy during the inactive portion of intermittent motions, but still coasts forward, so that intermittent motions are always energetically favorable.

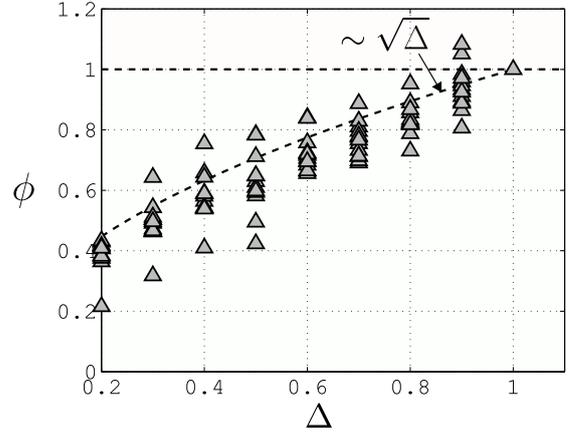

Figure 8: Ratio of energy expended by intermittent motions to energy expended by continuous motions as a function of duty cycle for $\theta_0 = 15°$, all frequencies, $m^* = 1$, and $A_w^* = 10$.

Apart from energy spent on swimming, aquatic animals also expend energy on metabolic processes. To capture this, we consider the metabolic energy ratio

$$\psi = \frac{E + E_m}{E_0 + E_{m0}},$$

where $E_m$ is the energy spent due to metabolic processes. We will assume that the mean power spent on metabolic processes is the same for continuous and intermittent motion, and that it is a constant fraction $c_m$ of the power lost in continuous swimming, $P_0$. Hence,

$$\psi = \frac{E + E_m}{E_0 + E_{m,0}}$$
$$= \frac{Pt + c_m P_0 t}{P_0 t_0 + c_m P_0 t_0} = \frac{(P + c_m P_0) d/U}{(1 + c_m) P_0 d_0 / U_0} = \frac{P + c_m P_0}{(1 + c_m) P_0} \frac{U_0}{U},$$

since $d = d_0$. Values of $\psi < 1$ indicate motions that are energetically favorable. The experimental results are plotted in figure 9. We have varied the metabolic power fraction $c_m$ from 0 to 2, typical in biology (Di Santo and Lauder, private communication). With $c_m = 0$ (darkest symbols), the plot is the same as in figure 8, indicating that intermittent motions are energetically favorable. As the metabolic power fraction is increased, however, the trend reverses. For large enough values of $c_m$ (approximately bounded by $c_m > 1$), intermittent motions expend more energy than continuous motions, a trend that increases as the duty cycle decreases.

Based on the mean thrust, power, and speed scalings, the metabolic energy ratio should scale as

$$\psi = \frac{P + c_m P_0}{(1+c_m)P_0} \frac{U_0}{U} \sim \frac{(\Delta + c_m)P_0}{(1+c_m)P_0} \frac{1}{\sqrt{\Delta}} = \frac{\Delta + c_m}{(1+c_m)\sqrt{\Delta}}.$$

From figure 9 we see that the data and the scaling tell the same story: even though intermittent motions expend less energy on swimming, the metabolic losses can play a significant role because intermittent motions take more time to traverse a given distance than continuous motions. The extra time to travel will increase the metabolic energy losses, and this effect may dominate the benefits gained in swimming energy losses.

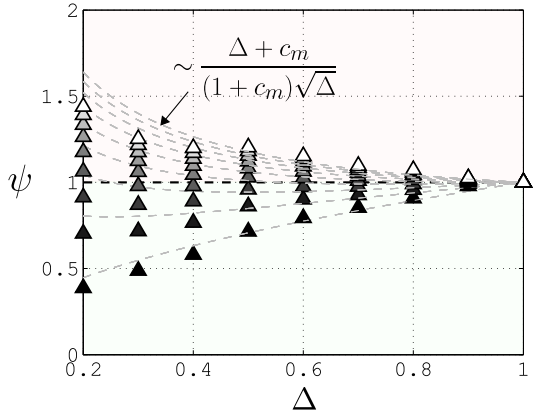

Figure 9: Ratio of energy expended by intermittent motions to energy expended by continuous motions, including metabolic energy losses, as a function of duty cycle for $\theta_0 = 15°$, all frequencies, $m^* = 1$, and $A_w^* = 10$. Each point is an average over all frequencies. The color denotes the value of the metabolic power fraction $c_m$, 0 to 2 in intervals of 0.25 (dark to light).

Another comparison we can make is to consider the energy ratio $\phi$, but restrict ourselves to motions which produce the same mean speed, that is, $\phi|_{U_{\text{mean}}}$. For example, it may be that a continuous motion produces the same mean speed as an intermittent motion with a duty cycle of 0.5 actuating at twice the frequency, but which motion expends more energy? The answer to this question will reveal which gait is best if a swimmer wants to traverse a given distance in a given amount of time. The results are plotted in figure 10. Interestingly, it appears that intermittent motions continue to be energetically favorable with the added time restriction. Energetically optimal duty cycles exist, and savings are greater for lower speeds, at least for the data considered here. Despite having to increase the frequency of actuation in order to match the mean speed of continuous motions, intermittent motions are nevertheless energetically favorable in this context.

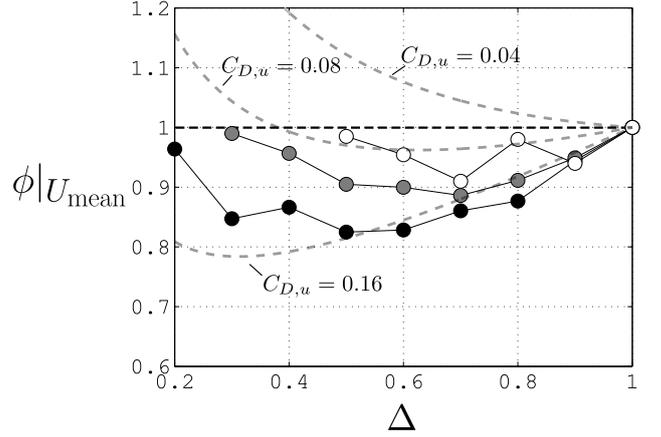

Figure 10: Ratio of energy expended by intermittent motions to energy expended by continuous motions, restricted to equal mean speeds, as a function of duty cycle for $\theta_0 = 15°$. Dashed lines correspond to equation 7, with $m^* = 1$, $A_w^* = 10$. The symbol grey scale corresponds to three values of (dimensional) mean speed chosen, $U_{\text{mean}} = 0.2, 0.25, 0.3$ (dark to light). The frequency of the intermittent motion was chosen so that it would have a speed equal to the continuous motion.

We can understand this behavior by using the scaling relations obtained here and in previous work [12]. The thrust coefficient approximately follows

$$\frac{C_T}{\Delta} \sim c_1 S t^2 - C_{D,u},$$

where $C_{D,u}$ is the offset of the thrust curve. Similarly, the power coefficient approximately follows

$$\frac{C_P}{\Delta} \sim c_2 S t^3.$$

Note that this relationship is different than given in [12], but it fits our data almost as well and is more convenient for our purposes here. Solving for the power gives

$$P \sim \Delta \cdot 4c_2 f^3 A^3 \rho s c = \Delta \cdot 4c_2 \left(\frac{T}{\Delta \cdot 2c_1 \rho s c} + \frac{C_{D,u} U^2}{4c_1}\right)^{3/2} \rho s c,$$

As shown earlier, the mean speed can be estimated well by equating the mean thrust of the propulsor to the mean drag of the body, so that

$$T = D = \frac{1}{2}\rho U^2 A_w C_D.$$

The power then becomes

$$P \sim \Delta \frac{c_2 U^3}{2 c_1^{3/2}} \left( \frac{1}{\Delta} \frac{A_w}{sc} C_D + C_{D,u} \right)^{3/2} \rho s c.$$

In this case, the energy ratio between intermittent and continuous motions is

$$\phi|_{U_{\text{mean}}} = \frac{E}{E_0} = \frac{Pt}{P_0 t_0} = \frac{P}{P_0},$$

since $t = t_0$. Hence,

$$\phi|_{U_{\text{mean}}} \sim \frac{\Delta \cdot \left( \frac{1}{\Delta} A_w^* C_D + C_{D,u} \right)^{3/2}}{\left( A_w^* C_D + C_{D,u} \right)^{3/2}}. \quad (7)$$

For the cases analyzed here, we chose $A_w^* = 10$ and $C_D = 0.01$. This expression is plotted in figure 10 for $C_{D,u} = \{0.04, 0.08, 0.16\}$ (for our data, $C_{D,u} = 0.08$). Note that the energy ratio is quite sensitive to the value of $C_{D,u}$. The expression is in qualitative agreement with the data, having a U-shape and indicating that there may be a particular duty cycle which is optimal.

## 5 Conclusions

The forces and energetics of intermittent swimming motions, characterized by the duty cycle, were analyzed and compared to continuous swimming. Water tunnel experiments on a nominally two-dimensional rigid foil pitching about its leading edge showed that mean thrust and power increased with increasing duty cycle, all the way up to continuous motion. The mean thrust and power data for different duty cycles were collapsed by dividing the mean thrust and power by the duty cycle, indicating that thrust and power in one burst cycle are not affected by the previous burst cycle. PIV measurements of the wake showed that the dominant structures shed into the wake were always counter-rotating pairs of vortices, regardless of the duty cycle. The PIV measurements corroborated the assertion that individual cycles of activity are unaffected by previous cycles.

Free swimming speed and energetics were analyzed by numerically integrating the measured experimental data. Although the experimental data was acquired for a foil moving at a constant speed, previous work showed that the thrust produced by a pitching foil is independent of speed so we expect that the thrust measured in stationary experiments would be the same as what would be measured in free swimming experiments. For a large range of area and mass ratios, the mean speed was then found to be the same as what would be calculated by assuming a constant speed and equating mean thrust with mean drag. The mean speed was lower than this steady speed only for parameters unlikely to be encountered in nature, indicating that the constant steady speed is a good approximation to the mean speed.

The energetics of intermittent motions were then compared to continuous motions according to three criteria: (i) energy expended in traversing a given distance; (ii) energy expended in traversing a given distance, including metabolic energy; and (iii) energy expended in traversing a given distance in a given time. Intermittent motions were generally energetically favorable as no energy is spent during the inactive portion of the motion where the swimmer still coasts forward. When metabolic energy losses were added, they could be high enough to make continuous swimming energetically advantageous.

The assertion that forces are unaffected by speed for pitching motions, shown in a previous study, was instrumental in being able to use the stationary experiments in the free swimming analysis. Although the forces produced by general motions (for example, combinations of pitch and heave) will depend on the speed, this work nonetheless highlights the potential benefits and pitfalls of intermittent motions as a swimming protocol.

This work was supported by ONR Grant N00014-14-1-0533 (Program Manager Robert Brizzolara). We would also like to thank Dr. Keith Moored for stimulating our interests in intermittent swimming.